\documentclass[conference]{IEEEtran}
\input epsf
\usepackage{hyperref}
\usepackage{tabularx}
\usepackage{graphicx}
\usepackage{xcolor}

\usepackage{etoolbox}
\makeatletter
\patchcmd{\@makecaption}
  {\scshape}
  {}
  {}
  {}
\makeatother
\begin{document}

\title{\LARGE A Novel Approach for Detection and Ranking of Trendy and Emerging Cyber Threat Events in Twitter Streams}

\author{\authorblockN{Avishek Bose, Vahid Behzadan, Carlos Aguirre, William H. Hsu\\ abose@ksu.edu, behzadan@ksu.edu, caguirre97@ksu.edu,  bhsu@ksu.edu}
\authorblockA{\authorrefmark{1}Department of Computer Science, Kansas State University, Manhattan, Kansas, 66506, USA}}



%


\maketitle

\begin{abstract}
We present a new machine learning and text information extraction approach to detection of cyber threat events in Twitter that are \textit{novel} (previously non-extant) and \textit{developing} (marked by significance with respect to similarity with a previously detected event). While some existing approaches to event detection measure novelty and trendiness, typically as independent criteria and occasionally as a holistic measure, this work focuses on detecting both novel and developing events using an unsupervised machine learning approach. Furthermore, our proposed approach enables the ranking of cyber threat events based on an importance score by extracting the tweet terms that are characterized as named entities, keywords, or both.  We also impute influence to users in order to assign a weighted score to noun phrases in proportion to user influence and the corresponding event scores for named entities and keywords. To evaluate the performance of our proposed approach, we measure the efficiency and detection error rate for events over a specified time interval, relative to human annotator ground truth.
\end{abstract}
\IEEEoverridecommandlockouts
\IEEEpeerreviewmaketitle

\section{Introduction}
This paper presents a new methodology for recognizing potential cyber threats using passive filtering and ranking in social text streams, particularly Twitter streams.  \textit{Passive} monitoring here refers to collecting intelligence and solutions of different cyber threats from different platforms using only text corpora and lists of named entities or keywords (e.g., gazetteers) rather than direct background knowledge of threats.  Twitter is examined as a high-bandwidth platform where both actors from both sides of cyberdefense, such as attackers and security professionals, post cybersecurity-related messages \cite{CyberHunt}. The overall goal of this work is to analyze these messages collectively to attain actionable insights and collect intelligence on emergent cyber threat events. Detecting events from social media includes a) \textbf{novel} event detection, including first stories or tweets about previously non-extant topics; and \textit{developing} events (especially for bursty topics, but also for non-bursty topics for which volume and aggregate important build up gradually).  In this work, we treat novelty of events from the developing nature of events (emergence) as \textbf{orthogonal} properties. This allows novel events that have not yet attained trending status or viral propagation to be tracked, while still incorporating traditional trend detection methods.

Recent research includes some work on detecting both novel and developing events in Twitter streams (e.g., \cite{Mario}\cite{Ifrim}), especially where emergence is defined as trending. However, only a few studies have further focused on detecting cyber threat events in Twitter streams. 
Furthermore, we propose an approach to the ranking of events with regards to their significance. While such a ranking generally depends on both the application domain and user objectives, the relative importance to a general community of interest, such as the cybersecurity community, can be imputed based on pervasiveness, spread rate, and novelty. In this study, we also rank the two types of events based on the order of their corresponding importance score to show how much a particular event is important compared to proximate events within a user-specified time range of a reference tweet.  In contrast with large document corpora, analyzing short documents such as tweets presents some specific semantic challenges towards extracting terms, relationships, patterns, and actionable insights in general. For example, terms mentioned in a short tweet lack context, and there is less co-occurrence data in the entire corpus on which to base expressible relations between named entities or terms.  

Our system takes as input a user-specified maximum interval of detection for related cyber threat events, within an original tweet that is deemed relevant. The full text of this tweet, or quoted part of a retweet, is captured. Social network parameters such as indegree (number of followers) are calculated and normalized by range. The text bodies of tweets are  vectorized using term frequency-inverse document frequency (TFIDF) \cite{tfIdf}, the resulting TFIDF vectors are clustered using the \textit{DBSCAN} \cite{DBSCAN} density-based clustering algorithm, noise points are discarded, and the concatenated text contents of each cluster are ranked using the \textit{TextRank} algorithm \cite{TextRank}, to obtain representative keywords and named entities that represent \textbf{potential} events.  We then identify different scenarios: a) novel and developing story; b) novel story only; c) developing story only; d) not an event based on heuristics that are described in Section \ref{Sec:Method}.
Additionally, we also calculate an importance score for each event based on the heuristics presented in Section \ref{Sec:Method}. Finally, we tag each event according to their descriptive features and provide a rank based on their importance scores.

Key \textbf{novel contributions} of this work are as follows:
\begin{enumerate}
 \setlength{\itemsep}{-2ex}
 \setlength{\parskip}{0ex} 
 \setlength{\parsep}{0ex}
\item We detect both trendy and novel types of events related to cybersecurity from Twitter streams. \hfil\break 
\item We provide a method for the ranking of potential cyber threat events according to their importance score based on keywords, as well as their named entity confidence and user influence scores.\hfil\break
\item The proposed method can be tuned to capture important cybersecurity events based on user-specified parameters.  
\end{enumerate}


\section{Related Approaches}\label{sec:Related}
This section briefly summarizes key methodologies for cyber threat detection from text corpora, particularly social media.



Dabiri et al.\cite{Sina} analyzed traffic related tweets for detecting traffic event by applying deep learning models, including convolutional and recurrent neural networks incorporating a \texttt{word2vec}-based word embedding layer to represent terms. This approach performs well but is domain-dependent and highly costly in terms of manual annotation for high-throughput sources of training data such as Twitter. In contrast, \textit{TwitInfo} \cite{Adam} incorporates a new streaming algorithm that automatically discovers peaks of event-related tweets and labels them from the tweets’ texts. This approach, however, focuses only burstiness of tweets and ignores both user influence and novelty with respect to developing events.

Rupinder et al. \cite{Rupinder} also proposed a framework based on deep learning for extracting cyber threat and security-related insights from Twitter, categorizing three types of threats (examples of which are Distributed Denial of Service (DDoS) attacks, data breaches, and account hijacking). From text documents, events are extracted using a) target domain generation; b) dynamically-typed query expansion; and c) event extraction. This approach employs both syntactic and semantic analysis using dependency tree graphs and convolutional kernel, but is highly computationally intensive due to the cost of autoencoder training.

Sceller et al. \cite{Sceller} uses unsupervised learning to detect and categorize cybersecurity events by analyzing cybersecurity-related Twitter posts based on a set of seed keywords specified for each level taxonomy.  This algorithm is prone to false negatives because it may not detect potential cyber threat events as events in the first place.

Ranade et al. \cite{Ranade} propose a method for processing threat-related tweets using the Security Vulnerability Concept Extractor (SVCE) which generates tags about cybersecurity threat or vulnerabilities such as means of an attack, consequences of an attack, affected software, hardware, and vendors. This approach does not generalize to the user communities as it is personalized for individual users' system profiles.

Edouard \cite{Edouard} propose a framework that utilizes Named Entity Recognition (NER) and ontology reasoning (using \textit{DBPedia}), along with classification learning approaches such as Naive Bayes, SVM, and a deep neural network (Long Short Term Memory / Recurrent Neural Network, aka LSTM-RNN), for category tag imputation. The graph algorithm \textit{PageRank} is used to rank candidate items for information retrieval.

The approach of Lee et al. \cite{Lee} focuses on community communication and influence to detect cyber threats by grouping highly contributing Twitter users and scores them as an expert community to get information to be explored and then to be efficiently exploited. This framework incorporates four components: a) an interface to the Twitter social media platform; b) a flexible machine learning system interface for document categorization; c) a mixture-of-experts weighting and extraction scheme; d) a new topic detector.  This framework is highly dependent on expertise and data quality.

A method by Sapienza et al. \cite{Sapienza} considers various web data sources to generate indication of warnings for detecting upcoming potential cyber threats. While potentially extensible to named entites discoverable by set expansion, this approach is focused on detecting "novel words" and does not yet incorporate a full contextualized topic model, feature weighting model, or method of user influence.

Finally, the work of Alan et al. \cite{Alan} is based on a supervised learning approach to train an extractor for extracting new categories of cybersecurity events by seeding a small number of positive event example over a significant amount of unlabeled data. As with previous approaches, it does not yet incorporate full NER nor allow for entity set expansion.

\section{Background}\label{sec:Background}
This section presents a brief review of the key technologies that are adopted in our proposed framework for threat event detection in tweets.
\subsection{Named Entity Recognition (NER)} In general, NER is an information extraction task aimed at locating and classifying the names of specific entities such as persons, organizations and locations, based on analysis of text units such as n-grams and noun phrases. Generic entities such as numerical quantities are sometimes also included.  In our analysis, NER is used to discover the names of entities in reported cyber threats. Key objectives of using machine learning to improve NER are: a) set expansion to broaden the set of cyber threats based on synonymy and other relationships that can be inferred by text pattern analysis; b) feature weighting for relevance or salience; c)  relationships that are discoverable from data; d) confidence scoring.

\subsection{TextRank}
The TextRank algorithm \cite{TextRank} is an extended version of Google PageRank \cite{PageRank} algorithm that aims to determine keywords by generating a word graph from a given text document unlike determining high ranked webpages that is done by the PageRank algorithm.  The TextRank score calculates importance of a word from given a text that is identical to PageRank score works for webpages. The importance however associated with a vertex is determined based on the votes that are cast for it, and the vertices' score casting these votes.

\subsection{TFIDF}
TFIDF \cite{tfIdf} is a information retrieval method used in various purpose such as word co-occurence based document vectorization, word ranking, document similarity calculation, etc. In information retrieval, TF (term frequency) refers to term frequency of a particular word in a document, while IDF (inverse document freqency) refers to inverse document frequency of a word in the whole corpus of documents. 
\subsection{\textit{DBSCAN}}
\textit{DBSCAN} \cite{DBSCAN} is a density-based clustering approach works by enforcing a minimum number of data points (\emph{MinPts}) inside a specified-radius neighborhood (\emph{Eps}) of a data point to make a \textit{density-reachable} cluster; this process continues until no points on the frontier are density-reachable, then restarts with a new initial point.

\section{Proposed Method}\label{Sec:Method}
Analyzing Twitter texts for getting valuable insights has always been an issue because of its unstructured way of writing and the length of tweets. In this section we go through some steps described in the subsections below. 
\subsection{Tweet Collection and Early Annotation}In this analysis we used Twitter data collected through four days from $6^{th}$ September to $9^{th}$ September in 2018 and a small portion of data collected in $30^{th}$ and $31^{th}$ August, 2018 that is stored in MongoDB database.  As our main focus is to getting insights of cybersecurity-related events and rank their scores, we crawled Twitter data using the Twitter API based on some security related keywords. Without applying security related keywords, the crawled Twitter data would be generalized to all domains and thus the result would be biased to detecting general kind of events. This datasets was manually annotated whether the tweets are relevant to cyber security or not by taking help from four annotators for our earlier work \cite{CyberHunt}. Although we used security related keywords for crawling the Twitter data, many of those tweets are irrelevant or promotional. That is why the annotation plays a crucial role here and the resulting dataset of this process are available at \cite{data set}. In this study, we initially have 21368 tweets and working with the annotated data, we found 11111 tweets are related to cyber security. We apply our algorithm on those cyber security related tweets and the whole tweets' datasets individually. We took the full text of each tweet if the tweet is not quoted or retweeted from any other users. If any tweet is retweeted or quoted from any other user, we take the original retweeted or quoted tweet full text. Then, we let an user to give a numeric value input as the number of time intervals based on tweet occurrence. This process divides the whole time period of tweet occurrence into some equal time chunks based on the number that we are taking from the user. Thus, for each time interval a number of tweets are aggregated into a chunk based on their corresponding occurring time. 
\subsection{Tweet Pre-processing and Cleaning} As we stated earlier that tweet text is very unstructured that contains a lot of misspelled words and sometime the text is not a complete sentence. That is why we apply a tool named SymSpell \cite{SymSpell} to correct the misspelled words. Then we take only the characters from the tweet text that are alphanumeric and remove all punctuation characters. Then we remove all the stopwords from the text because these are so frequently occurred over the whole data set that may reduce the analysis performance. After that we consider cleaning the tweets' texts in both cases either by applying stemming or without stemming. We removed all the words or tokens which lengths are only one.
\subsection{Influential Twitter user  Impact} Influential users tweets are valuable for detecting important cyber security related events. That is why we keep records of each Twitter user with their number of followers corresponding to their posted cyber security related tweets. As the number of followers represents is directly proportional to the influence of the user, their used words in cyber security related tweets are also important. So, for each time interval, we normalize the values of follower numbers for each user using Min-Max normalization. This normalization process normalizes each value between 0 and 1. We then assign the normalized value of users' follower number to each of their used noun phrases in tweets. Here, we used python nltk library to extract noun phrases from tweets. If similar words are used by several tweet user, we keep the highest normalized value of an user for each word used in a tweet. Now, for each time interval, for all tweets, each noun phrase has a corresponding value that represents its weight inherited by its user. This value will also be used to calculate event score.

\subsection{Determining Algorithm Design Architecture}
In this study, we apply a very popular word vectorization method in NLP domain named tfIdf \cite{tfIdf} that is based on word co-occurrence in documents to make word vector for each tweet from the data set. After doing process mentioned above for all tweets in a time interval, it generates a tfIdf matrix. We found this method gives a better performance compared to the word semantic relation based approaches that are discussed later in Section VI. Then, we apply the DBSCAN \cite{DBSCAN} density based clustering algorithm using the aforementioned tfIdf matrix to find cluster of similar meaning tweets. These clusters can represent the potential events. However, we did not apply the K-means clustering algorithm because we did want to limit the number of events found in our analysis, for each time interval. For this analysis, we ignore the noise points generated by applying DBSCAN because in our observation over the data set we found that the noise points are conveying a very little impact to find cyber security related events.
\subsection{Event Detection Heuristics and Scoring}
\begin{figure}
\includegraphics[width=0.5\textwidth, height=0.5\textheight,keepaspectratio]{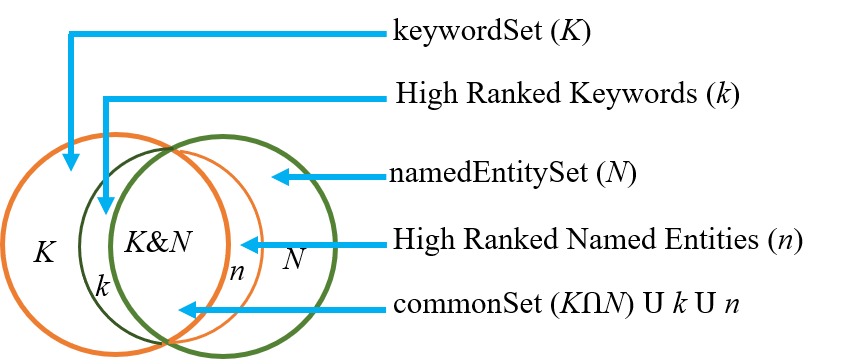}
\caption{Graphical representation of \emph{commonSet}, \emph{keywordSet} and \emph{namedEntitySet}}\label{Setop}
\end{figure}

Firsly,  we aggregate all tweet texts in a cluster into a single text. Then we simulate named entity and keyword identification process on the aggregated text by applying TextRazor \cite{TextRazor} online Named Entity recognition API and TextRank \cite{TextRank} algorithm from Gensim library respectively. Additionally, the TextRazor provides a Confidence score for each named entity and TextRank from Gensim \cite{Gensim} also provides a score for each keyword based on word graph mentioned earlier in Section IV. We apply two different set rules to determine the type of the events and and their corresponding score that will be used to rank the cyber security related events. To formulate our idea into implementation, we produce four different sets of token set mentioned below.
\begin{enumerate}

\item \emph{commonSet}-refers the set of words that are common in both named entity and keyword. Additionally, we also take some higher scored named entities and keywords from two sets \emph{namedEntitySet} and \emph{keywordSet} mentioned later respectively to add the tokens to the \emph{commonSet}. 
\item \emph{keywordSet}-refers the set of words that only appear in the keyword set and not common with the named entity set.
\item \emph{namedEntitySet}-refers the set of words that only appear in the named entity set and not common with the keyword set.
\item \emph{unionSet}-this set keeps all named entities and keywords. 
\end{enumerate}
The Figure \ref{Setop} clearly depicts the graphical illustration of four different token sets mentioned above. Here \emph{commonSet}, \emph{keywordSet}, \emph{namedEntitySet} and \emph{unionSet} are represented by $k\cup(K\cap N)\cup n$, \emph{K}-\emph{N}, \emph{N}-\emph{K} and $K \cup (K \cap N)\cup N$ respectively. Here, \emph{k} and \emph{n} are the set of highly scored keywords and named entities and can be represented as $\emph{k}\in \emph{K}$ and $\emph{n}\in \emph{N}$ respectively.
\subsubsection{Determine Event Novelty} We store all the tokens of the set \emph{namedEntitySet} and \emph{commonSet} into another set of tokens named as \emph{noveltyCheckerSet} for all clusters generated for all time intervals. We are storing all these tokens because we are checking the similarity of the tokens from the set of a subsequently generated cluster to the stored tokens' set \emph{noveltyCheckerSet} to determine whether the newly generated cluster has some novelty or not based on a cosine similarity threshold value determined empirically.
\subsubsection{Determining Trendiness} If the similarity score reaches the defined threshold \emph{cosineThresh}, we determine the working cluster is just trendy except the very first cluster because this cluster would not find any different set to compare the similarity. We also take a user defined threshold of number of tweets \emph{tweetThresh} to determine getting a trendy event. Thus, if the the number of tweets does not reach to the value mentioned by the user, it will be an unnoticeable event. However, if tweets from a cluster satisfies the cosine similarity threshold \emph{cosineThresh} as well as number of tweets threshold \emph{tweetThresh}, still it may not represent a noticeable event because it may be a spamming of banal topic. So, we apply a different heuristic if the length of the \emph{commonSet} set is greater than the one fifth(0.20) times of the \emph{namedEntitySet} set, only then the cluster will be counted as trendy. We are checking this because a big cluster of tweets' texts will have so many named entities that will mean a variety of topics but a single cluster should be biased towards a single topic described in it.
\subsubsection{Determining Novelty} Now, if the cosine similarity of stored token set of tokens \emph{noveltyCheckerSet} and working cluster token set is less than the threshold value \emph{cosineThresh}, the working cluster could be a potential novel event. However, we set a threshold value minimum three tweets to be in the cluster to refer it as an event. Now, if the number of tweets is greater or equal than the user defined threshold value of trendiness \emph{tweetThresh}, the cluster is addressed as ``Novel and Trendy" but if the number of tweets in the cluster is less than the number of user defined threshold value, it is addressed as ``Just Novel". That is how we determine a type of a generated cluster of tweets as an event type.
\subsubsection{Event Score Calculation Process} As we mentioned earlier the ranking of cyber threat related events is also an important task, we are motivated to calculate score of each event if an event finds a event type based on our empirically defined heuristics mentioned above. We calculate scores for each of defined events individually by applying different heuristics. As the confidence score of a named entity in the TextRazor and the score of a keyword in the TextRank algorithm are different in scale, we apply sigmoid function to normalize each scores of a named entity and keyword respectively. As every token is stored in a dictionary for each generated cluster, we update the score of each token if it considered both named entity and keyword by adding two scores after normalizing by the sigmoid function.
\subsubsection{Score Calculation for Trendy Event} Now, for a ``Just Trendy" event firstly we calculate the entity score of the event by adding the scores of each token included in the \emph{commonSet} and then multiplying the total added value with the value of total number of tweets that makes an event as trendy. Secondly, we added the value of each noun phrases corresponding to the event's tweets where the noun phrases are inherited from the value of influential users' followers. This score are then added to the initially calculated entity score mentioned above for the aggregated tweets' texts to get the total score.  We calculate the entity score like this mentioned above because, this will assign a higher score to a trendy event either if tweets in a same topic appear so many times in a cluster or number of tokens in the \emph{commonSet} is higher. This heuristic assumes that even if the event topic does not appear so many times in corresponding tweets compared to the other highly appeared event topics, because of the number of common tokens in both name entity and keywords, the heuristic give importance to those tokens as important tokens.
\subsubsection{Score Calculation for Trendy and Novel Event} Then for a ``Novel and Trendy" event, we added the values of all the tokens of the set being generated by the union operation of \emph{keywordSet} and \emph{commonSet}. Afterwards, we multiply the added value with the value of total number of tweets that represent the event to calculate the entity score. Then we again added the value of each noun phrases corresponding to the event's tweets where the noun phrases are inherited from the value of influential users' followers. This score are then added to the entity score like discussed earlier in the above paragraph to get the total score. We calculate the entity score like this because this event is already proved to be a novel event and we need to consider whether it has tokens common in both named entity and keyword set to check the main topic discussed about in this event. Additionally, we also need to consider the keywords appeared in this event to check which topics are also mentioned in the tweet texts of the novel event. That means a novel and trendy event will get a very higher score compared to the other events.
\subsubsection{Score Calculation for Just Novel Event} Now, for the ``First Story" event, we consider to keep a set of tokens generated by differentiating \emph{keywordSet} from \emph{unionSet} and then doing union operation with the \emph{commonSet}. This resulting set stores all the named entities with the keywords which have very high score. Then, we add all the values of the corresponding tokens in the resulting set and multiply the added value with the user defined threshold value \emph{tweetThresh} for being an event as trendy to get the entity score. Then we repeat the procedure of adding the value of noun phrases to the entity score of the working event. We are calculating the entity score like this, because if the event is just novel, it will not appear in so many tweets and that is how it may loose its importance. So, by means of giving importance to this kind of events, we are multiplying the total score of resulting set of tokens by the value of \emph{tweetThresh}. Thus, it can get at least as importance as any trendy event can acquire. We choose the aforementioned resulting set because, we need to consider the novelty of an event that is based on the the confidence score of the named entity and some high ranked keywords. There is no mean to consider the whole keyword set right now because in this case, it seems useless to other trendy topics discussed except the common ones with named entity set.
\subsubsection{Ranking Scores of Events} 
\begin{figure*}
\centering
\includegraphics[width=\textwidth,height=0.36\textheight,keepaspectratio]{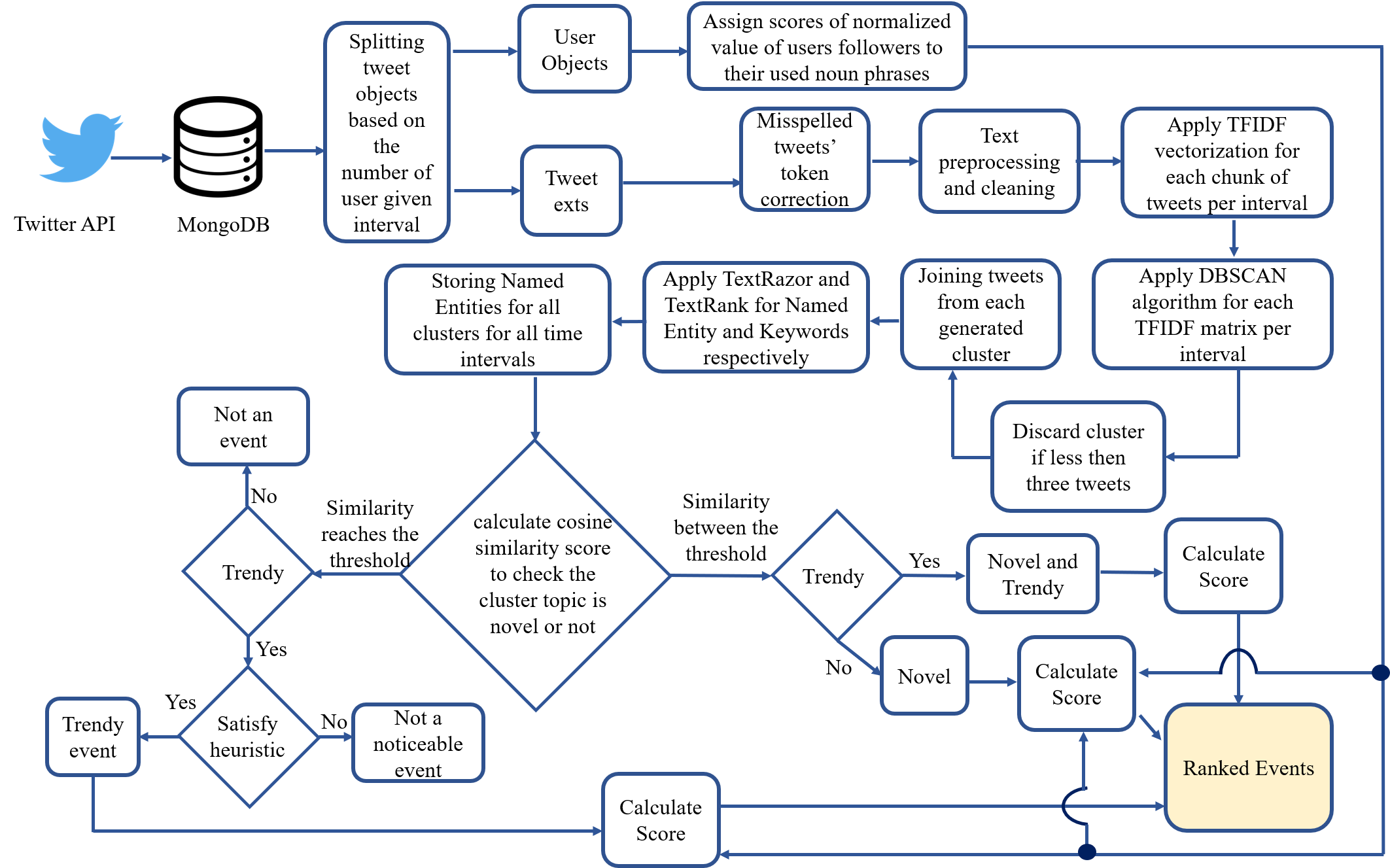}
\caption{Flowchart of the proposed approach}\label{Flowchart}
\end{figure*}

Our proposed approach repeats the above mentioned condition checking for each cluster whether to determine as an event or not and score calculation for each cluster that is only considered as an event for each time interval. Finally, we rank each event by ordering their total score for each time interval. The flow chart of our proposed approach is depicted in Figure \ref{Flowchart}.

\subsection{Annotation Approach}
\label{sec:AnnMethod}
To evaluate the performance of our proposed approach, we compared the results of our proposed method with a manually annotated list of events. A subset of 301 tweets collected in sequence in the window starting at 2018-08-30 23:00:08 CST to 2018-09-02 10:50:19 CST was manually annotated according to i. impact, ii. tweet count, and iii. worldwide effect to be considered as an event. We also consider three categories whether they are i. first story and novel, ii. trending or developing, and iii. novel and trending. For validation, we check the ratio of correctly detected events in that window to the total number of relevant events, and the ratio of correctly detected events to the total number of detected events. 

\section{Experimental Results}\label{sec:Results}

\subsection{Simulation}

\begin{figure*}
\centering
\includegraphics[width=\textwidth,height=0.3\textheight,keepaspectratio]{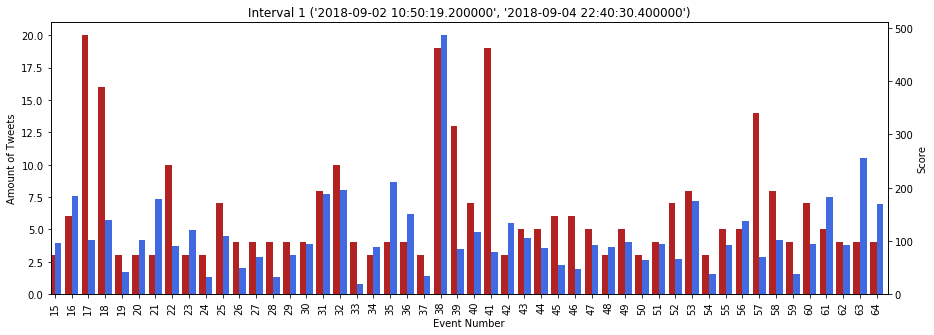}
\caption{Event plot of the second time interval proposed approach}\label{Plot}
\end{figure*}

For our analysis, we use scikit learn  \cite{scikit} to calculate tfIdf matrix \cite{tfIdf}, to apply DBSCAN \cite{DBSCAN} algorithm and to use cosine similarity. The parameters assignments for making the tfIdf matrix are \emph{max\_df} = 0.90, \emph{max\_features} = 200000, \emph{min\_df} = 0.01 and \emph{ngram\_range} = (1,1). Then, parameter assignments for DBSCAN algorithm are \emph{eps} = 1 and \emph{min\_samples} = 3. Again, we use the cosine similarity threshold as 0.5 for similarity checking for trendiness. Table \ref{tab:simResult} shows the result of five time intervals collectively from 2018-08-30 23:00:08 to 2018-09-02 10:50:19.200000, from 2018-09-02 10:50:19.200000 to 2018-09-04 22:40:30.400000, from 2018-09-04 22:40:30.400000 to 2018-09-07 10:30:41.600000, from 2018-09-07 10:30:41.600000 to 2018-09-09 22:20:52.800000 and from 2018-09-09 22:20:52.800000 to 2018-09-12 10:11:04 by Interval 1, 2, 3, 4 and 5 respectively. We keep only detected True Positive events and represent in Table \ref{tab:simResult}. From Table \ref{tab:simResult} we can see for the first interval (2018-08-30 23:00:08 to 2018-09-02 10:50:19.200000) we have total 145 tweets and total 15 events. Moreover, out of 15 events we got no ``Trendy Event", 1 ``Novel and Trendy Event" and 14 ``Novel Event". This description will be continued for rest of the intervals similarly. In Table \ref{tab:keyWord} we show the extracted keywords for each event for the first time interval. The keywords mentioned in the Table \ref{tab:keyWord} are used to detect cyber security related events for the first time interval. For better representation, We show only the plot of the $2^{nd}$ time interval in Fig. \ref{Plot} because of the paper space limitation. This figure depicts the found events in x axis, the amount of tweets on left side of figure \ref{Plot} in y axis and the event score on the right side  of figure \ref{Plot} in y axis. The red vertical bar represents number of tweets and the blue vertical bar represents event score for each detected event.

\begin{table}
\vskip5pt
\caption{Summery Result of five time intervals; NT:Number of Tweets; JT: Just Trendy; TN: Trendy and Novel; FS: First Story; TE: Total Number of Events}
\centerline{
\vbox{\offinterlineskip
\hrule
\halign{&\vrule#&
\strut\quad#\hfil\quad\cr
&Interval&&NT&&JT&&TN&&FS&&TE&\cr
height2pt&\omit&&\omit&&\omit&&\omit&&\omit&&\omit&\cr
\noalign{\hrule}
height2pt&\omit&&\omit&&\omit&&\omit&&\omit&&\omit&\cr
&1&&145&&0&&1&&14&&15&\cr
&2&&314&&0&&0&&50&&50&\cr
&3&&812&&1&&7&&37&&45&\cr
&4&&1239&&0&&9&&18&&27&\cr
&5&&297&&4&&0&&5&&11&\cr
height2pt&\omit&&\omit&&\omit&&\omit&&\omit&&\omit&\cr}
\hrule}}
\label{tab:simResult}
\end{table}

\begin{table*}
\vskip5pt
\caption{Summery result of time interval 1('2018-08-30 23:00:08', '2018-09-02 10:50:19.200000')}
\centerline{
\vbox{\offinterlineskip
\hrule
\halign{&\vrule#&
\strut\quad#\hfil\quad\cr
&Event NumberN&&Keywords&\cr
height2pt&\omit&&\omit&\cr
\noalign{\hrule}
height2pt&\omit&&\omit&\cr
&0&&{'security', 'android (operating system)', 'android', 'wi-fi', 'privacy'}&\cr
&1&&{'microsoft', 'disclosed', 'twitter', 'windows', 'microsoft windows', 'hacker discloses'}&\cr
&2&&{'website', 'catalonia', 'spain', 'banking', 'bank', 'inf'}&\cr
&3&&{'based', 'huff', 'buffer overflow'}&\cr
&4&&{'vulnerability (computing)', 'security', 'repository', 'critical vulnerability', 'apache', 'inf'}&\cr
&5&&{'vulnerability', 'resource consumption', 'prior', 'resource', 'rsa', 'bleach'}&\cr
&6&&{'task','windows','patch,'scheduler'}&\cr
&7&&{'security', 'android (operating system)', 'android', 'data', 'privacy', 'tracking'}&\cr
&8&&{'cracking ransom', 'coin', 'free', 'ransom', 'cybersex', 'net'}&\cr
&9&&{'vulnerability (computing)', 'patch', 'spyware', 'phishing', 'inf sec cube security', 'patched', 'malware'}&\cr
&10&&{'security', 'website'}&\cr
&11&&{'plus', 'pump'}&\cr
&12&&{'version', 'web server', 'debugger', 'skype', 'update', 'denial service', 'exploit (computer security)'}&\cr
&13&&{'cisco systems', 'service', 'cisco'}&\cr
&14&&{security', 'photo','service'}&\cr
height2pt&\omit&&\omit&\cr}
\hrule}}
\label{tab:keyWord}
\end{table*}

\begin{table*}
\vskip5pt
\caption{Summery result of time interval 1('2018-08-30 23:00:08', '2018-09-02 10:50:19.200000');EN: Event Number;EL: Event Link; TC:Tweets Count\newline NESR:Normalized Event Score Rank;ET: Event Type; AER: Annotator Event Ranking; DBR: Difference between Rankings; FS: First Story (novel); FST: First Story and Trendy (developing)}
\centerline{
\vbox{\offinterlineskip
\hrule
\halign{&\vrule#&
\strut\quad#\hfil\quad\cr
&EN&&TC&&Event Score&&NESR&&ET&&EL&&AER&&DBR&\cr
height2pt&\omit&&\omit&&\omit&&\omit&&\omit&&\omit&&\omit&&\omit&\cr
\noalign{\hrule}
height2pt&\omit&&\omit&&\omit&&\omit&&\omit&&\omit&&\omit&&\omit&\cr
&0&&5&&167.6084&&5&&FS&&\href{https://threatpost.com/android-os-api-breaking-flaw-offers-up-useful-wifi-data-to-bad-actors/137085/}{link1}&&4&&1&\cr
&1&&7&&211.033&&3&&FS&&\href{http://www.excel.blue/search////2018_09_28/exploit.html}{Link2}&&2&&1&\cr
&2&&21&&190.5950&&4&&FS&&\href{https://www.hackread.com/ddos-attack-anonymous-catalonia-cripples-bank-of-spain-website/}{Link3}&&3&&1&\cr
&3&&3&&55.3226&&10&&FS&&\href{}{NA}&&13&&3&\cr
&4&&12&&110.6048&&9&&FS&&\href{https://searchsecurity.techtarget.com/news/252447943/Another-patched-Apache-Struts-vulnerability-exploited}{Link4}&&7&&2&\cr
&5&&4&&130.4169&&8&&FS&&\href{https://nvd.nist.gov/vuln/detail/CVE-2016-0887}{Link5}&&8&&0&\cr
&6&&3&&17.2225 &&14&&FS&&\href{https://duo.com/decipher/windows-task-scheduler-flaw-has-temporary-fix}{Link6}&&10&&4&\cr
&7&&6&&145.7938&&7&&FS&&\href{https://www.zdnet.com/article/android-operating-system-vulnerability-leaks-device-data-allows-user-tracking/}{Link7}&&5&&2&\cr
&8&&8&&154.3115&&6&&FS&&\href{https://www.zdnet.com/article/cracking-ransomware-ransomwarrior-victims-can-now-retrieve-files-for-free/}{Link8}&&6&&0&\cr
&9&&5&&389.7082&&2&&FS&&\href{https://www.darkreading.com/risk/how-hackers-hit-printers-/d/d-id/1332715}{Link9}&&9&&7&\cr
&10&&4&&40.0639&&13&&FS&&\href{}{NA}&&14&&1&\cr
&11&&7&&46.4706&&12&&FS&&\href{https://healthitsecurity.com/news/critical-cybersecurity-vulnerability-found-in-bd-alaris-plus-pump}{Link10}&&12&&0&\cr
&12&&51&&391.3391&&1&&FST&&\href{https://borncity.com/win/2018/11/28/dos-vulnerability-in-microsoft-skype-for-business/}{Link11}&&1&&0&\cr
&13&&5&&52.8906&&11&&FS&&\href{https://www.cisco.com/c/en/us/support/docs/security/asa-5500-x-series-firewalls/212972-anyconnect-vpn-client-troubleshooting-gu.html}{Link12}&&11&&0&\cr
&14&&4&&16.9345&&15&&FS&&\href{}{NA}&&15&&0&\cr
height2pt&\omit&&\omit&&\omit&&\omit&&\omit&&\omit&&\omit&&\omit&&\omit&\cr}
\hrule}}
\label{tab:perform}
\end{table*}

\subsection{Annotation-Based Validation}

\subsubsection{Design Selection Approach}There are few decision we had to make to formulate the design architecture of this algorithm. Firstly, we thought to analyze the semantic relation between the words of each tweet text to get the insights of cyber security events. That is why we previously applied doc2Vec \cite{doc2vec} which is an extended application of word2vec \cite{word2vec} for getting similar meaning tweets to find events from the data sets but we could not get any satisfactory result because we found that shallow neural network model text domain tools like doc2Vec \cite{doc2vec} works based on word vector embeddings that does not perform well for short and noisy text data set. Embedding methods did not work properly in short texts because tokens in a short text have a thin contextual relation between each other and this relation get worse due to misspelled and incomprehensible tokens. A sample result of doc2Vec applying hypermeter values \emph{vector\_size} = 300, \emph{min\_count} = 2 and \emph{epochs} = 45 respectively is shown in Table \ref{tab:doc2VecRes} that exhibits most similar tweet and second most similar tweet of a particular tweet that does not have any noticeable similarity with any of those tweet document whereas a document must show similarity with at least to itself. Similarity score of each tweet to the particular  tweet is represented inside the parentheses in the first column. Again, we thought in a different way to apply LDA \cite{lda} to find some topics that may represent events. Since the tweet text is very short, almost all of the time a tweet does not represent more that one event. That is why, we decided to apply LDA \cite{lda} on the aggregated tweet texts from corresponding time intervals. This, approach also fails to show the expected result because of the incoherent nature of tweet text. Due to the space limitation of the we could not present in this work. Thus, we decided to apply very popular tfIdf vectorization because we found that words in a tweet text has a few semantic relation between each other and word co-occurrence is better option to apply in this domain. Consequently, we found a better performance by comparing the result with the previously applied approaches.

\begin{table}[htbp]
\caption{Sample result of doc2Vec}
\centering
\begin{tabular}{|p{0.4\linewidth}|p{0.5\linewidth}|}
\hline
\multicolumn{1}{|c|}{\textit{\textbf{Terms}}} & \multicolumn{1}{c|}{\textit{\textbf{Texts}}}                                                   \tabularnewline \hline
Document                                           & guides on fixing sql injections vulnerabilities sql injection technique exploits security vulnerability occurring database layer application the vulnerability present user input either                                                    \tabularnewline \hline
Most Similar (0.8027611970901489)                                              & free vps server ddos protected hosting                                                                                  \tabularnewline \hline
Second Most Similar (0.6457577347755432)                                          & cvnway just ddos server                                                                   \tabularnewline \hline
Median ( 0.21316465735435486)                                               & minibb bbfuncsearchphp table sql injection                                                   \tabularnewline \hline
Least (-0.4030833840370178)                                             & ransomware weapon used cyber attacks elixir ng news source trust                                                                    \tabularnewline \hline
\end{tabular}
\label{tab:doc2VecRes}   
\end{table}

\subsubsection{Validation of the Approach}
Table \ref{tab:confMatrix} shows the performance result of our proposed approach according to the evaluation methodology described in Section \ref{sec:AnnMethod}. The annotators annotate 301 and found total 20 events and 6 tweet clusters that are not events.  On the other hand, the our algorithm found total 16 events. Now, 15 events out of 16 events are real events (True Positive) included in 20 ground truth but one event is False positive. So, the True Positive, False Positive, False Negative and True Negative rates are 75\%, 16.67\%, 25\% and 83.33\% respectively and we got a good precision value that is 93.75\%. An interesting news is that we can only stream a very small amount of tweets per millisecond approximately 1\% of the total tweet posted that is addressed in this web article \cite{tweetLink}. So, the Twitter data itself only is not sufficient to detect all of the ongoing cyber security event and that is why we limit ourself to calculate recall score by keeping track published cyber security events in online. In Table \ref{tab:perform}, we present the fifteen true positive events, along with their tweet count and their corresponding scores. We order the events by their corresponding scores and match with the annotators' annotations. The $4^{th}$ and the $7^{th}$ column of the Table \ref{tab:perform} present proposed approach event ranking and annotators' ranking respectively and comparing the $4^{th}$ and the $7^{th}$ column of the table, we can see the annotators predictions are quite similar to our approach in case of event detection and event ranking. The validity of the detection approach can be checked by clicking the link mentioned in the $6^{th}$ column to see the reports published in authentic blogs and newspapers. The $5^{th}$ column represents the type of events detected by our algorithm. The sum squared error (SSE) of the event ranking of our approach and annotator's ranking is 86 by calculating the difference mentioned in the $8^{th}$ column of the table. 
\begin{table}[htbp]
\caption{Confusion matrix of the algorithm's generated result}
\centering
\begin{tabular}{|p{0.25\linewidth}|p{0.2\linewidth}|p{0.2\linewidth}|}
\hline
\multicolumn{1}{|c|}{\textit{\textbf{Total Population}}} & \multicolumn{1}{c|}{\textit{\textbf{Ground Truth positive}}} & \multicolumn{1}{c|}{\textit{\textbf{Ground Truth negative}}}                                                   \tabularnewline \hline
Derived positive                     & True Positive=75\%                      & False Positive=16.67\%                                                     \tabularnewline \hline
Derived negative       & False Negative=25\%                                       & True Negative=83.33\%                                                                                  \tabularnewline \hline
\end{tabular}
\label{tab:confMatrix}   
\end{table}

\section{Conclusion}\label{sec:Conclusion}
We presented a novel machine learning and text information extraction method for the detection of cyber threat events from tweets. We considered two types of such events, those that are novel, and those that are further developments of previously detected tweets. Furthermore, we proposed an approach for the ranking of cyber threat events based on an importance score computed based on the named entities and keywords in the text of tweets.  We also impute influence to users in order to assign a weighted score to noun phrases in proportion to user influence and the corresponding event scores for named entities and keywords. To evaluate the performance of our proposals, we measure the efficiency and detection error rate for events over a specified time interval, relative to human annotator ground truth, and demonstrate the feasibility of its application in detecting cyber threat events from tweets. Future directions of this research include the extension of our current method for detection and ranking of sub-events in each cyber threat event. Moreover, the heuristics applied in this work are presented as proofs of concept, while leaving room for further enhancement and customization per user requirements. As further venue of future word, the methodology used for influence measurement of users can be extended via means such as meta-network modeling and link extraction of the dynamic social network of users that are active in the cybersecurity domain. 


\end{document}